\documentclass[pra,aps,twocolumn,showpacs]{revtex4}

\usepackage{amssymb}
\usepackage{graphicx}

\usepackage{amsmath}

\def\<{\langle}
\def\>{\rangle}
\def\ket#1{|#1\>}
\def\bra#1{\<#1|}

\begin{document}

\title{Efficient experimental estimation of fidelity of linear optical quantum Toffoli gate}

\author{M. Mi\v{c}uda}
\affiliation{Department of Optics, Palack\'{y} University, 17. listopadu 1192/12, CZ-771 46 Olomouc, Czech Republic}

\author{M. Sedl\'{a}k}
\affiliation{Department of Optics, Palack\'{y} University, 17. listopadu 1192/12, CZ-771 46 Olomouc, Czech Republic}

\author{I. Straka}
\affiliation{Department of Optics, Palack\'{y} University, 17. listopadu 1192/12, CZ-771 46 Olomouc, Czech Republic}

\author{M. Mikov\'{a}}
\affiliation{Department of Optics, Palack\'{y} University, 17. listopadu 1192/12, CZ-771 46 Olomouc, Czech Republic}

\author{M. Du\v{s}ek}
\affiliation{Department of Optics, Palack\'{y} University, 17. listopadu 1192/12, CZ-771 46 Olomouc, Czech Republic}

\author{M. Je\v{z}ek}
\affiliation{Department of Optics, Palack\'{y} University, 17. listopadu 1192/12, CZ-771 46 Olomouc, Czech Republic}

\author{J. Fiur\'{a}\v{s}ek}
\affiliation{Department of Optics, Palack\'{y} University, 17. listopadu 1192/12, CZ-771 46 Olomouc, Czech Republic}

\begin{abstract}
We propose an efficiently measurable lower bound on quantum process fidelity of $N$-qubit controlled-Z gates. This bound is determined by average output state fidelities for $N$
partially conjugate product bases.  A distinct advantage of our approach is that only fidelities with product states need to be measured while
keeping the total number of measurements much smaller than what is necessary for full quantum process tomography.
As an application, we use this method to experimentally estimate quantum process fidelity $F$ of a three-qubit linear optical quantum Toffoli gate and we find that $F\geq 0.83$.
We also demonstrate the entangling capability of the gate by preparing  GHZ-type three-qubit entangled states from input product states.
\end{abstract}


\maketitle

As the complexity of quantum information processing devices increases, there is a growing demand for reliable
and efficient methods of their characterization. Traditionally, the experimentally implemented quantum operations are being characterized by quantum
process tomography which provides their full description \cite{Nielsen00,Paris04}. However, a complete quantum tomography requires
resources which grow exponentially with the number of qubits. In recent years, increasing attention has been therefore paid to methods such as compressed sensing \cite{Gross10,Shabani11}
 or Monte Carlo sampling \cite{Flammia11,Silva11,Steffen12},  that allow to reduce the overhead of quantum process characterization.
 This is possible if we make some a-priori assumption about the structure
 of the reconstructed quantum object or if we do not seek a complete tomographic description and we are satisfied instead with determination of a particular parameter
 such as fidelity \cite{Emerson07,Bendersky08,Koch13}.

 In 2005, it was shown by Hofmann that the quantum process fidelity can be estimated by measuring the
 average quantum state fidelities $F_1$ and $F_2$ for only two conjugate bases \cite{Hofmann05}. The quantum process fidelity $F_\chi$ is then lower bounded
 according to
 \begin{equation}
 F_\chi \geq F_1+F_2-1.
 \label{FHofmann}
 \end{equation}
  This procedure has received a considerable attention
  and it was utilized in several experiments to estimate the fidelity of a quantum CNOT gate \cite{Okamoto05,Bao07,Clark09,Gao10,Gao10b,Zhou11}.
  In this case,
  it is sufficient to measure the CNOT truth table in the computational basis and in the dual basis obtained from the computational basis
  by single-qubit Hadamard transformations. In this dual basis, the gate also acts as a CNOT but the roles of control and target qubits are reversed.
  The great practical advantage of this characterization of a CNOT gate is that all the required input probe states as well as all the
  corresponding output states whose fidelity should be measured are product states. Therefore, the fidelities can be directly determined
  by single-qubit measurements. The fidelity bound (\ref{FHofmann}) is applicable to arbitrary multiqubit quantum gates.
  However, going beyond the two-qubit CNOT gate, it would typically require measurements of fidelities of output states that are entangled
  \cite{Lanyon11}.

  In this paper, we propose a generalization of the Hofmann fidelity bound to multiqubit controlled-Z gates which preserves all the experimentally
  desired features present for the two-qubit CNOT gate. In particular, product multiqubit probe states are considered and
  only fidelities of product states have to be measured. Furthermore, the total required number of measurement settings  is only $N2^N$, while at least
   $2^{3N}$ measurement settings are necessary for a full quantum process tomography of an $N$-qubit gate specified by $2^{4N}-2^{2N}$ parameters 
   for deterministic gates or $2^{4N}$ parameters for probabilistic gates.
  Here a measurement setting represents a specific combination of input state preparation and an output measurement.
  As an application, we employ this method to experimentally estimate quantum process fidelity of a three-qubit linear optical controlled-Z gate (CCZ gate) that is equivalent to the
  Toffoli gate  up to single-qubit Hadamard transform on a target qubit.
  The Toffoli gate is a crucial part of many quantum information processing schemes \cite{Nielsen00} and it represents one of the most complex quantum circuits experimentally implemented to date
\cite{Lanyon09,Monz09}. A peculiar feature of linear optical quantum gates is that they are probabilistic \cite{Kok07}. In practice, it may happen due to various imperfections
  that the success probability of the gate depends on the input state and is not a constant. We shall show that the fidelity bounds hold even
  in such case but the average state fidelities must be calculated as weighted averages, with the weights given by relative probabilities of success.

  Let $|0\rangle$ and $|1\rangle$ denote the computational basis states of a qubit.
  The CCZ gate flips the sign if and only if all three qubits are in state $|1\rangle$,
  \begin{equation}
  U_{\mathrm{CCZ}}= \mathbb{I} -2|111\rangle\langle 111|,
  \end{equation}
  where $\mathbb{I}$ denotes the identity operator. In the computational basis, we explicitly have $U_{\mathrm{CCZ}}|abc\rangle=(-1)^{abc}|abc\rangle$.
  A Hadamard transform $H$ on a target qubit converts the sign flip of CCZ gate into the bit flip of Toffoli gate.
  Recall that $H|0\rangle=|+\rangle$ and $H|1\rangle=|-\rangle$, where $|\pm\rangle=\frac{1}{\sqrt{2}}(|0\rangle\pm|1\rangle)$.

  According to the Choi-Jamiolkowski isomorphism \cite{Jamiolkowski72,Choi75}, any quantum operation $\mathcal{E}$ can be represented by a positive semidefinite
  operator $\chi$ on a tensor product of input and output Hilbert spaces. Let $|\Phi_3^{+}\rangle=\sum_{a,b,c=0}^{1}|abc\rangle_{\mathrm{in}}|abc\rangle_\mathrm{out}$
  denote a maximally entangled state on two copies of a three-qubit Hilbert space. The Choi matrix of operation $\mathcal{E}$ can be determined by applying this operation
  to one part of the maximally entangled state, $\chi = \mathcal{I}_{\mathrm{in}}\otimes \mathcal{E}_\mathrm{out}(\Phi_3^{+})$,
  where $\Phi_3^{+}=|\Phi_3^{+}\rangle\langle \Phi_3^{+}|$ is a short-hand notation for a density matrix of a pure state.
  For any input density matrix $\rho_{\mathrm{in}}$, the corresponding output density matrix $\rho_{\mathrm{out}}=\mathcal{E}( \rho_{\mathrm{in}})$
   can be calculated as
  $\rho_{\mathrm{out}}=\mathrm{Tr}_{\mathrm{in}}[\rho_{\mathrm{in}}^T\otimes \mathbb{I}_{\mathrm{out}} \,\chi]$, where $\mathrm{Tr}_{\mathrm{in}}$ denotes partial
  trace and $T$ stands for transposition. We shall consider general probabilistic operations
  and $\mathrm{Tr}[\rho_{\mathrm{out}}]=\mathrm{Tr}[\rho_{\mathrm{in}}^T\otimes \mathbb{I}_{\mathrm{out}} \,\chi]$
  then represents the success probability for the input $\rho_{\mathrm{in}}$.
  In particular, the Choi matrix of a unitary CCZ gate reads
  \begin{equation}
  \chi_{\mathrm{CCZ}}= (\mathbb{I}\otimes U_{\mathrm{CCZ}}) \,\Phi_3^{+}  (\mathbb{I}\otimes U_{\mathrm{CCZ}}^\dagger).
  \end{equation}
  The process fidelity of a quantum gate
   $\chi$ with respect to the ideal CCZ gate can be defined as a normalized overlap of the Choi matrices,
  $F_\chi=\mathrm{Tr}[\chi \,\chi_{\mathrm{CCZ}}]/( 8\mathrm{Tr}[\chi] )$,
  where the factor $8=\mathrm{Tr}[\chi_{\mathrm{CCZ}}]$  accounts for the normalization of $\chi_{\mathrm{CCZ}}$.

  \begin{table}[b]
\caption{List of the input three-qubit product states $|\psi_{j,k}\rangle$.}
\begin{ruledtabular}
\begin{tabular}{ccccccc}
$j$ &   $k=1$ & $k=2$ & $k=3$ & $k=3'$   \\ \hline
 1 & $|\! + \! 00\rangle$ &  $|0\! + \! 0\rangle$ &   $|00 +  \rangle$ & $|\!+\!+0 \rangle$   \\
 2 & $|\! + \! 01\rangle$ &  $|0\! + \! 1\rangle$ &   $|00 -  \rangle$ & $|\!+\!+ 1  \rangle$  \\
 3 & $|\! + \! 10\rangle$ &  $|0\! - \! 0\rangle$ &   $|01 +  \rangle$ & $|\!+\!- 0  \rangle$  \\
 4 & $|\! + \! 11\rangle$ &  $|0\! - \! 1\rangle$ &   $|01 -  \rangle$ & $|\!+\!- 1  \rangle$  \\
 5 & $|\! - \! 00\rangle$ &  $|1\! + \! 0\rangle$ &   $|10 +  \rangle$ & $|\!-\!+ 0  \rangle$  \\
 6 & $|\! - \! 01\rangle$ &  $|1\! + \! 1\rangle$ &   $|10 - \rangle$  & $|\!-\!+ 1  \rangle$  \\
 7 & $|\! - \! 10\rangle$ &  $|1\! - \! 0\rangle$ &   $|11 +  \rangle$ & $|\!-\!- 0  \rangle$  \\
 8 & $|\! - \! 11\rangle$ &  $|1\! - \! 1\rangle$ &   $|11 -  \rangle$ & $|\!-\!- 1  \rangle$  \\
 \end{tabular}
\end{ruledtabular}
\end{table}

  Our goal is to lower bound the gate fidelity by suitably chosen average state fidelities. Motivated by
  the symmetry of the CCZ gate which is invariant with respect to the permutation of the qubits, we propose to characterize the CCZ gate
  by measuring the average output state fidelities for three complementary product bases where two of the qubits are prepared in the computational
  basis states $|0\rangle,|1\rangle$ while the third qubit is prepared in the Hadamard basis states $|+\rangle,|-\rangle$. The probe product states
  $|\psi_{j,k}\rangle$ are specified in Table I for all three bases $k=1,2,3$. It can be easily checked that all the corresponding
  output states $|\psi_{j,k}^{(\mathrm{out})}\rangle=U_{\mathrm{CCZ}}|\psi_{j,k}\rangle$ are also product states and the two bases $\left\{|\psi_{j,k}\rangle\right\}_{j=1}^8$
  and $\{|\psi_{j,k}^{(\mathrm{out})}\rangle\}_{j=1}^8$ coincide.
  Let $\rho_{j,k}=\mathrm{Tr}_{\mathrm{in}}[\psi_{j,k}^T\otimes \mathbb{I}_{\mathrm{out}} \chi]$
  denote the (unnormalized) output state for the input $|\psi_{j,k}\rangle$.
  The fidelity of this output state with the ideal output can be expressed as
  \begin{equation}
  f_{j,k}= \frac{\langle \psi_{j,k}^{(\mathrm{out})} |  \rho_{j,k}|\psi_{j,k}^{(\mathrm{out})}\rangle}{\mathrm{Tr}[\rho_{j,k}]}=
  \frac{1}{p_{j,k}}\mathrm{Tr}[\psi_{j,k}^T \otimes \psi_{j,k}^{(\mathrm{out})} \, \chi].
  \end{equation}
 Here $p_{j,k}=\mathrm{Tr}[\psi_{j,k}^T \otimes \mathbb{I}_{\mathrm{out}} \, \chi] $ denotes the probability of success of the gate for input
 $|\psi_{j,k}\rangle $ and $\psi_{j,k}=|\psi_{j,k}\rangle\langle \psi_{j,k}|$ denotes a density matrix of a pure state $|\psi_{j,k}\rangle$.
 We define the average state fidelity for $k$th basis as a weighted mean of $f_{j,k}$ with weights equal
 to the success probabilities $p_{j,k}$ \cite{Bell12},
 \begin{equation}
 F_k= \frac{\sum_{j=1}^8 p_{j,k}f_{j,k}}{\sum_{j=1}^8 p_{j,k}}.
 \label{Fmeandefinition}
 \end{equation}
 Since $\sum_{j=1}^8 \psi_{j,k}=\mathbb{I}$ for all $k$, it holds that $\sum_{j=1}^8 p_{j,k}=\mathrm{Tr}[\chi]$
 and we can express the mean fidelities in a compact matrix form $F_k=\mathrm{Tr}[R_k \chi]/\mathrm{Tr}[\chi]$, where
  \begin{equation}
  R_k=\sum_{j=1}^{8} \psi_{j,k}^T\otimes U_{\mathrm{CCZ}}\psi_{j,k}U_{\mathrm{CCZ}}^\dagger.
  \end{equation}

 The gate fidelity $F_{\chi}$ can be lower bounded by the three above defined average state fidelities as follows,
  \begin{equation}
  F_{\chi} \geq F_1+F_2+F_3-2,
  \label{Fbound}
  \end{equation}
  which generalizes the Hofmann bound (\ref{FHofmann}) and is valid for both deterministic and probabilistic quantum operations.
  In order to prove this bound, we rewrite it as
  \begin{equation}
  F_{\chi}-F_1-F_2-F_3+2 =\frac{\mathrm{Tr}[R\chi]}{\mathrm{Tr}[\chi]} \geq 0,
  \label{Finequality}
  \end{equation}
  where $R= \frac{1}{8}\chi_{CCZ} -R_1-R_2-R_3+2\mathbb{I}.$
  It can be shown by explicit calculation that the matrix $R$ is positive semidefinite, which immediately implies the inequality (\ref{Finequality}).
  Note that $R \geq 0$ holds irrespective of the actual form of the unitary $U_{\mathrm{CCZ}}$.
  Thus the fidelity bound (\ref{Fbound}) is actually completely general and it holds for all three-qubit unitary operations.
  The output states $|\psi_{j,k}^{\mathrm{(out)}}\rangle$ will be product states for all unitaries $U$ that are diagonal in the computational basis \cite{SI}.
  However, for other gates the output states may be entangled.
  We stress that it is important to calculate $F_k$ as wieghted means (\ref{Fmeandefinition}) because if $F_k$
  would be calculated as ordinary means, $F_k=\sum_{j=1}^8 f_{j,k}$/8, then the bounds (\ref{FHofmann}) and (\ref{Fbound})
  could actually overestimate the gate fidelity for certain trace-decreasing operations.
  The fidelity bound (\ref{Fbound}) can be generalized to $N$-qubit gates and it can be shown that
  \begin{equation}
  F_\chi \geq \sum_{k=1}^N F_k -N+1,
  \label{FboundNmain}
  \end{equation}
  where $F_k$ is the average fidelity for input product basis states where all qubits are prepared in the computational basis states
  except for  the $k$th qubit which is prepared in the Hadamard basis states.
  A detailed analytical proof of the bound (\ref{FboundNmain}) can be found in the Appendix \cite{SI}.
The bound will be tight and equality will hold in Eq. (\ref{FboundNmain}) if $\chi$ is either the desired unitary $U$, the unitary $U$ preceded by a $\pi$ phase flip on a single qubit, or 
 any mixture or coherent superposition of these unitary operations \cite{SI}.

\begin{figure*}[!t!]
  \centerline{\includegraphics[width=0.785\linewidth]{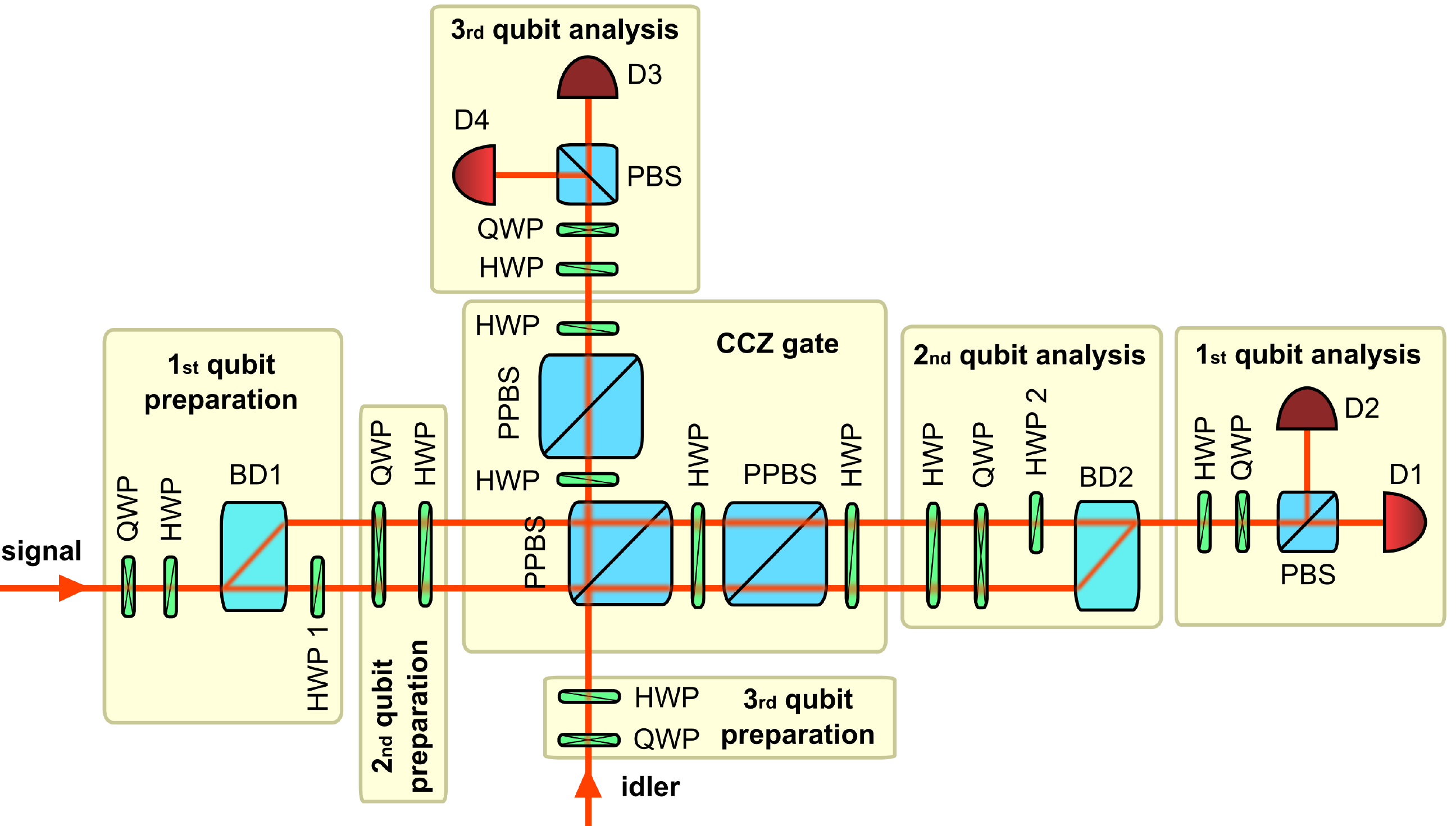}}
  \caption{Experimental setup. BD - calcite beam displacer, PPBS - partially polarizing beam splitter, PBS - polarizing beam splitter,
  HWP - half-wave plate, QWP - quarter-wave pate, D - single-photon detector.}
\end{figure*}

The experimental setup is shown in Fig. 1.
Orthogonally polarized time-correlated photon pairs with central wavelength of $810$ nm are generated in the process of
spontaneous parametric downconversion in a $2$ mm thick BBO crystal pumped by a CW laser diode with $75$ mW pump power \cite{Jezek11}.
The first qubit is encoded into spatial degree of freedom of the signal photon while the second and third qubits are encoded into
polarization of the signal and idler photons, respectively. The polarization states are
prepared and analyzed using half- and quarter-waveplates and polarizing beam splitters. The calcite beam displacer BD1 transforms
the input polarization state of the signal photon into state of spatial qubit and the beam displacer BD2 maps the spatial qubit back onto polarization.
The pair of beam displacers forms an inherently stable Mach-Zehnder interferometer. The HWPs which address only one path in the interferometer disentangle
input polarization and spatial qubits of the signal photon and ensure correct signal collection by BD2.

The CCZ gate is implemented by two-photon interference on a partially polarizing beam splitter PPBS with intensity transmittances
$T_H=1$ and $T_V=1/3$ for horizontally and vertically polarized photons, respectively \cite{Ralph02,Hofmann02,Okamoto05,Langford05,Kiesel05}.
The scheme also requires two additional PPBSs for balancing the amplitudes.
The $\pi$ phase shift due to two-photon interference on the central PPBS
occurs only if the signal photon travels through the lower interferometer arm  and both photons
are vertically polarized (logical qubit states $|1\rangle$).
This probabilistic CCZ gate operates in the coincidence basis \cite{Kok07} and its successful operation is indicated by detection of
 two-photon coincidences D1\&D3, D1\&D4, D2\&D3, or D2\&D4 at the output. Other detection events correspond to gate failure and are rejected and not used in subsequent analysis.
Ideal success probability of the CCZ gate is $1/9$ which is the maximum value achievable without the use of ancilla photons \cite{Lemr11}.

\begin{figure*}[!t!]
 \centerline{\includegraphics[width=0.95\linewidth]{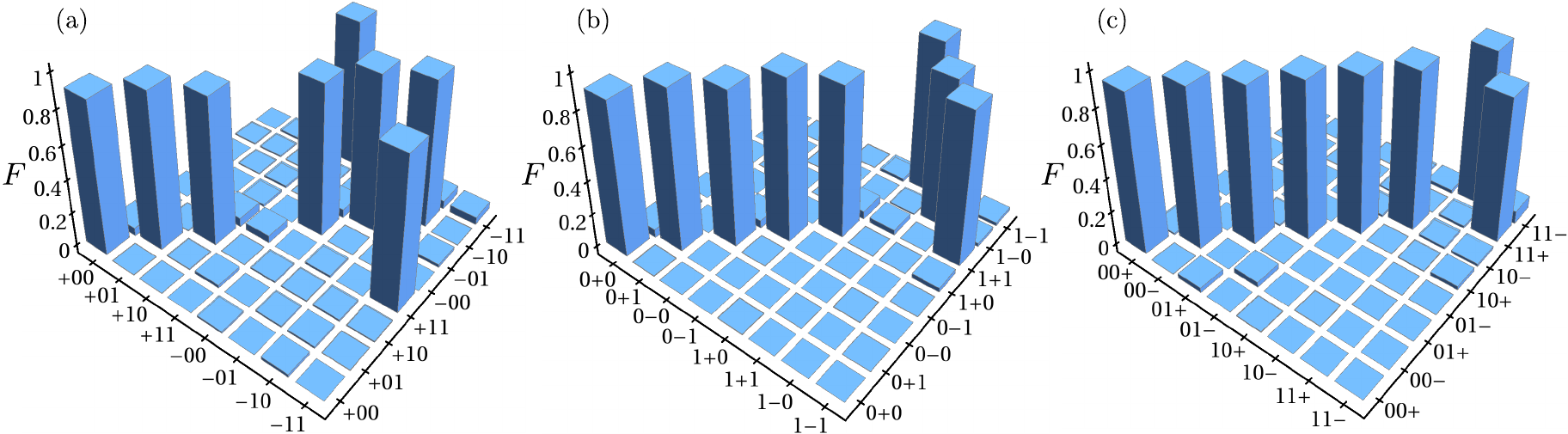}}
  \caption{Experimentally determined truth tables characterizing performance of the CCZ gate in three partially conjugate product bases $k=1$ (a), $k=2$ (b) and $k=3$ (c).}
\end{figure*}

We have prepared the input states $|\psi_{j,k}\rangle$, $k=1,2,3$, as listed in Table I and for each input we have
performed projective measurement on the output in the product basis $\left\{|\psi_{j,k}\rangle\right\}_{j=1}^8$.
The results are shown in Fig. 2 and they can be also interpreted as the computational basis truth tables of the Toffoli gates where
the target qubit is the first, the second, and the third qubit, respectively.
The parameters $f_{j,k}$ and $p_{j,k}$ necessary for evaluation of the average state fidelities $F_k$ were determined as follows.
Let $C_{j,j'}^{k}$ denote the number of detected coincidences corresponding  to projection onto $U_{\mathrm{CCZ}}|\psi_{j',k}\rangle$ for input $|\psi_{j,k}\rangle$.
Then $f_{j,k}=C_{j,j}^k/S_j^k$, where $S_{j}^k=\sum_{j'=1}^8 C_{j,j'}^k$ is the total number of coincidences for a given input. 
In our measurement, $S_j^k \approx 6.6\times 10^4$ on average.
Since all $C_{j,j'}^k$ were measured for the same time interval of $100$~s, the relative success probabilities can be determined 
as normalized total number of coincidences, $p_{j,k}=8 S_j^k/S^k$ where $S^k=\sum_{j=1}^8 S_j^k$.
The estimated $p_{j,k}$ lie in the interval $[0.902,1.079]$. On inserting the expressions for $f_{j,k}$ and $p_{j,k}$ into Eq. (\ref{Fmeandefinition}) we get
$F_k=\sum _{j=1}^8 C_{j,j}^k/S^k$,  which yields 
\begin{equation}
F_1= 0.928(1),\qquad
F_2=0.947(1), \qquad
F_3=0.955(1).
\label{Fkexp}
\end{equation}
The statistical errors 
represent $3$ standard deviations $\sigma_k$, which were determined assuming Poissonian statistics of the measured coincidences,
$\sigma_k^2=F_k(1-F_k)/S^k$.
If we plug the fidelities (\ref{Fkexp}) into formula (\ref{Fbound}) we obtain a lower bound on the fidelity of the implemented linear optical CCZ gate,
\begin{equation}
F_\chi \geq 0.830(2).
\label{Fest}
\end{equation}

\begin{figure}[!b!]
\centerline{\includegraphics[width=0.99\linewidth]{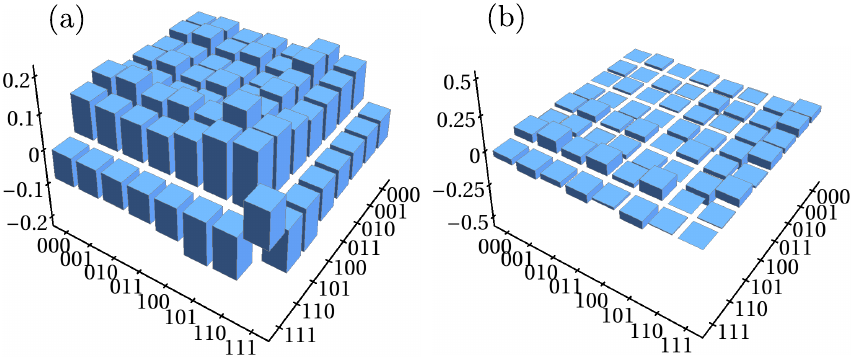}}
 \caption{Real (a) and imaginary (b) parts of reconstructed density matrix of a three-qubit state generated by the CCZ gate from input product state $|\!+\!++\rangle$.}
\end{figure}

For comparison, we have also experimentally determined the original Hofmann bound on the process fidelity (\ref{FHofmann}). For this purpose we consider
two bases labeled $k=3$ and $k=3'$ connected by Hadamard transforms of all three qubits, $|\psi_{j,3'}\rangle=H\otimes H\otimes H|\psi_{j,3}\rangle$.
The input states $|\psi_{j,3'}\rangle$ are explicitly listed in the last column of Table I. It holds that $F_\chi \geq F_3+F_{3'}-1$, where
 both fidelities are given by Eq. (\ref{Fmeandefinition}). This inequality can
 be equivalently expressed as $\mathrm{Tr}[R'\chi]/\mathrm{Tr}[\chi]\geq 0$, where $R'=\frac{1}{8}\chi_{\mathrm{CCZ}}-R_3-R_{3'}+\mathbb{I}$.
Similarly as before, one can explicitly show that $R'\geq 0$, which proves the above inequalities. The determination of $F_{3'}$ requires measurement
of fidelities of entangled output states, since for instance $U_{\mathrm{CCZ}}|\!+\!+1\rangle=\frac{1}{\sqrt{2}}(|0+\rangle+|1-\rangle)|1\rangle$.
Fortunately, in our setup the projection onto the maximally entangled states of qubits 1 and 2 can be accomplished deterministically
because both these qubits are carried by a single photon. We only have to rotate the HWP2 by $45^{\circ}$ which acts as a CNOT gate between the
polarization and spatial qubits carried by a single photon and transforms the maximally entangled states onto product states. From the measurements we
have determined $F_{3'}=0.921(1)$ which together with $F_{3}=0.955(1)$ as given in Eq. (\ref{Fkexp}) yields $F_\chi \geq 0.876(2)$. We can see that at the cost of measuring fidelities
of entangled states we obtain a slightly higher fidelity bound. This is universally valid because $R-R' \geq 0$.
Since $R_k-\frac{1}{8}\chi_{\mathrm{CCZ}} \geq 0$ for all $k=1,2,3,3'$, the average state fidelities also provide an upper bound on
the gate fidelity \cite{Hofmann05}, $F_\chi \leq \mathrm{min}(F_k)$. Specifically, we get $F_\chi \leq 0.921(1)$.

Finally, we explicitly demonstrate the capability of the three-qubit CCZ gate to generate entanglement from input separable states $|\psi_1\rangle|\psi_2\rangle|\psi_3\rangle$,
where $|\psi_j\rangle=c_0^j|0\rangle+c_1^j|1\rangle$ (partially entangled inputs are considered in the Appendix.
The corresponding output state reads
$ U_{\mathrm{CCZ}}|\psi_1\psi_2\psi_3\rangle= |\psi_1\psi_2\psi_3\rangle-2c_1^1c_1^2c_1^3 |111\rangle,$ hence the gate generates GHZ-type three-qubit entangled states \cite{GHZ89,Dur00}.
As an example, in Fig. 3 we plot the reconstructed output state corresponding to the input product state $|+++\rangle$.
The fidelity of the reconstructed state with the ideal state is $F=0.951$ and its purity $\mathcal{P}=\mathrm{Tr}[\rho^2]$
 reads $\mathcal{P}=0.959$.

In summary, we have proposed a lower bound on the quantum process fidelity of multiqubit quantum gates which generalizes
the original Hofmann bound and we used it to experimentally characterize linear optical quantum CCZ gate.
The advantage of our generalized fidelity bound is that only fidelities of product multiqubit quantum states need to be measured
and the required number of measurement settings is much smaller than what is needed for
full quantum process tomography or gate fidelity estimation by Monte Carlo sampling \cite{SI}.
We hope that our work will represent a useful addition to the toolbox of methods for
efficient and experimentally feasible characterization of quantum devices.

 \acknowledgments
 This work was supported by the Czech Science Foundation (13-20319S). M.S.
 acknowledges support by the Operational Program Education for Competitiveness - European Social Fund
 (project No. CZ.1.07/2.3.00/30.0004) of the Ministry of Education, Youth and Sports of the Czech Republic.

\newpage

\appendix*
\section{}

\subsection{Proof of the fidelity bound for $N$-qubit gates}
Our aim is to prove a bound on quantum process fidelity $F_{\chi}$ between the actually implemented quantum operation $\chi$ and an ideal target unitary gate $U$,
\begin{equation}
F_\chi \geq \sum_{k=1}^N F_k -N+1,
\label{FboundN}
\end{equation}
for arbitrary $N$-qubit unitary gate $U$.  Here $F_k$ denote average output state fidelities for $N$ different input product bases.
Analogously to the case of three qubit CCZ gate we define
\begin{equation}
|\Phi^{+}_N\rangle=\sum_{j_1,\ldots,j_N=0}^{1}|j_1,\ldots,j_N\rangle_{\mathrm{in}}|j_1,\ldots,j_N\rangle_\mathrm{out}
\label{PhiNplus}
\end{equation}
a maximally entangled state on two copies of an $N$-qubit input Hilbert space. The Choi matrix of an $N$-qubit unitary gate $U$ reads
\begin{equation}
  \chi_{\mathrm{U}}= (\mathbb{I}\otimes U) \,\Phi^{+}_N  (\mathbb{I}\otimes U^\dagger).
\end{equation}
We define the process fidelity of a quantum gate
   $\chi$ with respect to the $N$-qubit unitary gate $U$ as a normalized overlap of the Choi matrices,
  $F_\chi=\mathrm{Tr}[\chi \,\chi_{\mathrm{U}}]/( 2^N \mathrm{Tr}[\chi] )$,
  where the term $2^N=\mathrm{Tr}[\chi_{\mathrm{U}}]$ appears due to the normalization of $\chi_{\mathrm{U}}$.

The probe product states $|\psi_{j,k}\rangle$
are straightforward generalization of the states specified in Table I of the main paper.
In this case index $j=1,\ldots,2^N$ specifies an element of an $N$-qubit orthonormal product basis formed from basis vectors $\{\ket{0},\ket{1}\}$ on each qubit 
except for $k$-th qubit, where basis $\{\ket{+},\ket{-}\}$ is used. Explicitly, we can write
\begin{equation}
|\psi_{j,k}\rangle=\mathbb{I}^{\otimes k-1}\otimes H \otimes \mathbb{I}^{\otimes N-k}|j_1,\ldots,j_N\rangle,
\end{equation}
where $|j_1,\ldots,j_N\rangle$ is a computational basis state with $j_1 \ldots j_N$ forming  
a binary representation of integer $j-1$, and $H$ denotes the single-qubit Hadamard gate.
Note that an $N$-qubit controlled-Z gate $U_{\mathrm{C^NZ}}$ transforms the input product states $|\psi_{j,k}\rangle$ onto product states at the output and the preparation basis $\{|\psi_{j,k}\rangle\}_{j=1}^{2^N}$
and the measurement basis  $\{|\psi_{j,k}^{(\mathrm{out})}\rangle\}_{j=1}^{2^N}$ coincide.
The gate $U_{\mathrm{C^NZ}}$ is diagonal in computational basis and it introduces a $\pi$ phase shift if and only if all $N$ qubits are in state $|1\rangle$,
\begin{equation}
U_{\mathrm{C^NZ}}|j_1,\ldots,j_N\rangle= e^{i\pi \prod_{k=1}^Nj_k}|j_1,\ldots,j_N\rangle.
\label{UCNZ}
\end{equation}

Similarly to the three-qubit case, the average output state fidelity $F_k$ is defined as
\begin{equation}
F_{k}=\frac{\sum_{j=1}^{2^N} p_{j,k}f_{j,k}}{\sum_{j=1}^{2^N} p_{j,k}},
\end{equation}
where
\begin{equation}
p_{j,k}=\mathrm{Tr}[\psi_{j,k}^T \otimes \mathbb{I}_{\mathrm{out}} \, \chi],
\end{equation}
and
\begin{equation}
f_{j,k}=\frac{1}{p_{j,k}}\mathrm{Tr}[\psi_{j,k}^T \otimes \psi_{j,k}^{(\mathrm{out})} \, \chi]
\end{equation}
denote the success probability and output fidelity for input state $|\psi_{j,k}\rangle$, and $\psi_{j,k}=|\psi_{j,k}\rangle \langle \psi_{j,k}|$.
We can express $F_k$ in a matrix form $F_k=\mathrm{Tr}[R_k \chi]/\mathrm{Tr}[\chi]$, where
  \begin{equation}
  R_k=\sum_{j=1}^{2^N} \psi_{j,k}^T\otimes U \psi_{j,k} U^\dagger.
  \end{equation}
We can now rewrite the inequality (\ref{FboundN}) as:
 \begin{equation}
  F_{\chi}-\sum_{k=1}^N F_k + N-1 =\frac{\mathrm{Tr}[R \chi]}{\mathrm{Tr}[\chi]} \geq 0,
  \label{FinequalityN}
  \end{equation}
where
\begin{equation}
R= \frac{1}{2^N}\chi_{\mathrm{U}} - \sum_{k=1}^N R_k +(N-1)\mathbb{I}.
\end{equation}
Since the trace of a product of two positive semidefinite operators is non-negative, we would prove our claim by showing $R \geq 0$.
In order to show this we define an operator with the same eigenvalues $T \equiv (\mathbb{I}\otimes U^\dagger) R (\mathbb{I} \otimes U)$
and express it using the definition of $R$:
\begin{equation}
\label{TNop1}
T=(N-1)\mathbb{I}+\frac{1}{2^N}\Phi^{+}_N - \sum_{k=1}^N  \sum_{j=1}^{2^N} \psi_{j,k}^T\otimes \psi_{j,k},
\end{equation}
It is useful to divide the $2N$ qubit Hilbert space on which operator $T$ acts into two-qubit subsystems
formed by the $l$-th qubit of the input and the $l$-th qubit of the output. We introduce a unitary operator $W$ which groups together the $l$-th
input and output qubits,
\begin{equation}
W|j_1\ldots,j_N\rangle|k_1,\ldots,k_N\rangle=|j_1,k_1\rangle\ldots|j_N,k_N\rangle.
\end{equation}
In this way the maximally entangled state can be seen as
 $W\ket{\Phi^{+}_N}=2^{N/2} \ket{\Phi^{+}}_1 \cdots \ket{\Phi^{+}}_N$, where $\ket{\Phi^{\pm}}=\frac{1}{\sqrt{2}}(\ket{00}\pm\ket{11})$
 are the Bell states and the subscripts indicate the two-qubit subsystems. Let us now analyze terms of the type $\psi_{j,k}^T\otimes \psi_{j,k}$.
 For any $l$ they are factorized with respect to a subsystem of the $l$-th input and output qubit, because the state $\ket{\psi_{j,k}}$ is completely factorized.
 In any such subsystem we can find only one of four states $\ket{00}\bra{00}, \ket{11}\bra{11},\ket{++}\bra{++}, \ket{--}\bra{--}$. Let us note that
\begin{align}
\ket{00}\bra{00}+\ket{11}\bra{11}&= \Phi^{+} + \Phi^{-}  \nonumber \\
\ket{++}\bra{++}+\ket{--}\bra{--}&= \Phi^{+} + \Psi^{+},
\label{psisums}
\end{align}
where $\ket{\Psi^{\pm}}=\frac{1}{\sqrt{2}}(\ket{01}\pm\ket{10})$ are the other two Bell states.
 The identities (\ref{psisums}) allow us to rewrite the sums of projectors $\psi_{j,k}^T\otimes \psi_{j,k}$ as
\begin{eqnarray}
\label{psumid}
& & \!\!\!\! \!\!\!\!  \sum_{j=1}^{2^N} W \left(\psi_{j,k}^T\otimes \psi_{j,k}\right) W^\dagger \nonumber \\
& & = (\Phi^{+} + \Phi^{-})^{\otimes k-1}  \otimes (\Phi^{+} + \Psi^{+}) \otimes (\Phi^{+} + \Phi^{-})^{\otimes N-k}.  \nonumber \\
\end{eqnarray}
Using the four Bell states we can express the identity on the two qubit Hilbert space as $\mathbb{I}=\Phi^{+} + \Phi^{-} + \Psi^{+} + \Psi^{-}$.
This enables us to rewrite Eq. (\ref{TNop1}) using Eq. (\ref{psumid}) in the following form
\begin{widetext}
\begin{equation}
\label{TNop2}
\tilde{T}= (N-1)(\Phi^{+} + \Phi^{-} + \Psi^{+} + \Psi^{-})^{\otimes N}+ \Phi^{+ \otimes N}
- \sum_{k=1}^N  (\Phi^{+} + \Phi^{-})^{\otimes k-1}  \otimes (\Phi^{+} + \Psi^{+}) \otimes (\Phi^{+} + \Phi^{-})^{\otimes N-k}.
\end{equation}
\end{widetext}
where $\tilde{T}=WTW^\dagger$.
Since $W$ is unitary, operator $\tilde{T}$ has the same eigenvalues as $T$. Moreover,
$\tilde{T}$ is diagonal in the basis formed by tensor products of Bell states, hence the eigenvalues can be directly determined from the expression (\ref{TNop2}).
One eigenstate is given by tensor product of $N$ copies of $|\Phi^{+}\rangle$
and its eigenvalue reads $0$. $N 2^{N-1}$ eigenstates are formed by tensor products of a single copy of $|\Psi^{+}\rangle$ and $N-1$
copies of $|\Phi^{+}\rangle$ or $|\Phi^{-}\rangle$ and they all correspond to eigenvalue $N-2$.
For $k=1,\cdots,N-1$ we have ${N \choose k}$ eigenvalues formed by tensor products of $k$ copies of $|\Phi^{+}\rangle$ and $N-k$ copies of $|\Phi^{-}\rangle$ whose eigenvalue is equal to $N-k-1$.
All the remaining eigenstates correspond to the eigenvalue $N-1$. We can see that all the eigenvalues of $\tilde{T}$ are nonnegative for $N \geq 2$ which proves that $T$ is a positive semidefinite operator
 and as a consequence the same holds for operator $R$. This proves the fidelity bound (\ref{FboundN}).

In a similar fashion we can prove an upper bound on quantum gate fidelity,
\begin{equation}
F_\chi \leq F_k.
\label{Fchiupper}
\end{equation}
Taking into account Eq. (\ref{psumid}) and the definition of $F_\chi$, we find after some algebra that the inequality (\ref{Fchiupper}) is equivalent to the condition
\begin{widetext}
\begin{equation}
(\Phi^{+} + \Phi^{-})^{\otimes k-1}  \otimes (\Phi^{+} + \Psi^{+}) \otimes (\Phi^{+} + \Phi^{-})^{\otimes N-k-1} -\Phi^{+ \otimes N} \geq 0.
\end{equation}
\end{widetext}

This latter inequality is obviously satisfied which proves (\ref{Fchiupper}).

The bound (\ref{FboundN}) will be tight for any  $\chi$ whose support lies in the subspace spanned by the $N+1+4\delta_{N,2}$ eigenstates of $R$
with eigenvalue $0$. The multiplicity of the eigenvalue $0$ and the structure of the corresponding eigenstates follows from the above analysis of eigenstates and eigenvalues of operator  $\tilde{T}$.
In particular, for $N\geq 3$ this $N+1$ dimensional subspace is spanned by the mutually orthogonal basis states  $\mathbb{I}\otimes V_m |\Phi^{+}_N \rangle$ 
generated by $N+1$ unitary operations $V_m$, where $V_0=U$, and $V_m=U\Sigma_m$, $m=1,\ldots,N$. Here $\Sigma_m$ is a unitary operation that applies $\pi$ phase shift to $m$th qubit,
\begin{equation}
\Sigma_m= \mathbb{I}^{\otimes m-1}\otimes \sigma_Z \otimes \mathbb{I}^{\otimes N-m},
\end{equation}
and $\sigma_{Z}=|0\rangle\langle 0|-|1\rangle\langle 1|$. It is easy to check that if the actually implemented operation is $V_m$ then equality holds in Eq. (\ref{FboundN}) and the lower bound is thus
equal to the true gate fidelity $F_\chi$.  For $m=0$ we obviously have $F_\chi=1$ as well as $F_k=1,$ $k=1,\ldots,N$. On the other hand, if $m>0$
then $F_\chi=0$ and $F_k=1-\delta_{m,k}$, hence $\sum_{k=1}^N F_k=N-1$. Note that $F_{m}=0$  in this case because the phase flip 
$\sigma_Z$ maps
the input states $|+\rangle,\,|-\rangle$ of $m$th qubit onto orthogonal states. In contrast, all probe states $|\psi_{j,k\neq m}\rangle$ belonging to other bases are unchanged by $\Sigma_m$.
The fidelity bound (\ref{FboundN}) is thus tight if the actually implemented operation is either the desired unitary $U$ or a unitary $U$ preceded by a $\pi$ phase flip on a single qubit. 
The bound will be tight also for any mixture or coherent superposition of these operations.
The case $N=2$ is special because we have $4$ additional eigenstates with eigenvalue $0$. 
These $4$ eigenstates correspond to two-qubit unitary
operations $\sigma_X\otimes \mathbb{I}$, $\mathbb{I} \otimes \sigma_X$, $\sigma_X\otimes \sigma_Z$, and $\sigma_Z \otimes \sigma_X$, where $\sigma_X=|0\rangle\langle 1|+|1\rangle\langle 0|$.

Since each average state fidelity $F_k$ is larger or equal to the gate fidelity $F_\chi$,
the difference between the true fidelity and the lower bound (\ref{FboundN}) will increase at most linearly with the number of qubits N,
\begin{equation}
F_{\chi}-\sum_{k=1}^N F_k+N-1 \leq (N-1)\left(1-F_\chi\right).
\end{equation}
For large number of qubits and high level of noise it may happen that the bound (\ref{FboundN}) will be too small to be of any use.
However, with increasing $N$ the $N$-qubit controlled Z gate becomes very similar to the identity operation as wittnessed by their fidelity
\begin{equation}
F_{\mathrm{I}}=\frac{1}{2^{2N}}\left|\langle \Phi_N^{+}|\mathbb{I}\otimes U_{\mathrm{C^NZ}}|\Phi_N^{+}\rangle\right|^2.
\label{FIdef}
\end{equation}
On inserting the explicit expressions (\ref{PhiNplus}) and (\ref{UCNZ}) for $|\Phi_N^{+}\rangle$ and $U_{\mathrm{C^NZ}}$ into Eq. (\ref{FIdef}), we obtain
\begin{equation}
F_{\mathrm{I}}=1-2^{2-N}+ 2^{2-2N}.
\label{FIformula}
\end{equation}
In any experiment attempting to certify the quality of $N$-qubit controlled Z gate by fidelity measurement, fidelity $F_{\mathrm{I}}$ achievable by simply doing nothing has to be exceeded.
This requires an exponentially small gate infidelity, $1-F_\chi< 2^{2-N}(1-2^{-N})$. It is instructive to rewrite the fidelity bound (\ref{FboundN}) as
a bound on the gate infidelity $1-F_\chi$,
\begin{equation}
1-F_\chi \leq \sum_{k=1}^N (1-F_{k}).
\end{equation}
Since $F_k \geq F_{\chi}$, the bound (\ref{FboundN}) will certainly exceed $F_{\mathrm{I}}$ provided that the true gate infidelity will satisfy
$1-F_\chi< 2^{2-N}(1-2^{-N})/N$, which represents only a small extra overhead in gate quality compared to the dominant exponential factor $2^{-N}$.

Our construction of the set of input states $|\psi_{j,k}\rangle$ was tailored for the multiqubit controlled Z gates, but it is actually
well suited for a much wider class of all unitary operations $U_D$ that are diagonal in the computational basis,
\begin{equation}
U_D|j_1,\ldots ,j_N\rangle= e^{i\phi(j)} |j_1,\ldots ,j_N\rangle,
\label{UDdef}
\end{equation}
where $\phi(j)$ denotes an input-state dependent phase shift. For all unitary operations (\ref{UDdef}) it holds that the output states $U_D|\psi_{j,k}\rangle$ are product states 
of $N$ single-qubit states. Therefore, the output-state fidelities can be directly determined by $N$ single-qubit measurements similarly to the case of the controlled-Z gate.

However, for other $N$-qubit gates than (\ref{UDdef}) the output states $U|\psi_{j,k}\rangle$ may be entangled. In such case direct determination of output state fidelity would require measurement
in basis of entangled states, which may be difficult or even impossible for certain experimental implementations.
In that case one would have to estimate the fidelity of output entangled states from measurements in various product bases,
e.g. by the Monte Carlo sampling \cite{Flammia11,Silva11}. This would inevitably increase the required number of measurement settings. Consequently, other approaches such as Monte Carlo
sampling of the gate fidelity or measurement of the original Hofmann bound \cite{Hofmann05} may become preferable for such gates.

\subsection{Comparison with Monte Carlo sampling}

Our method of characterization of the $N$-qubit controlled-Z gates as well as the Monte Carlo sampling technique proposed in Refs. \cite{Flammia11,Silva11}
are designed such that it suffices to prepare product $N$-qubit input states and perform single-qubit measurements on the output.
This is important, because in many experiments it may be difficult or impossible to prepare entangled input states or to perform measurements
in entangled basis. The main advantage of our procedure is that it requires much smaller number of different
measurement settings than Monte Carlo sampling. A single measurement setting represents a specific combination of input state and output measurement.
In our experiment, both input states and measurement bases are set by waveplate rotations and each change of
measurement setting requires about $15$ seconds mainly due to a limited speed of motorized rotation stages.
By contrast, the data acquisition for a fixed setting is fast, because the average observed two-photon coincidence rates
are of the order of $80$ per second for each pair of detectors.

Experimental Monte Carlo estimation of fidelity of a three-qubit quantum Toffoli gate was experimentally demonstrated by Steffen \emph{et al.} \cite{Steffen12}. 
As shown in Ref. \cite{Steffen12}, fidelity of $3$-qubit Toffoli gate can be expressed
as a linear combination of average values of $232$ tensor products of single-qubit Pauli operators $\sigma_X$, $\sigma_Y$, $\sigma_Z$, and
$\sigma_I\equiv\mathbb{I}$.  Each tensor product includes six Pauli operators: $3$ for the input qubits and $3$ for the output qubits.
Experimental determination of a single average value requires sequential preparation of $2^3=8$ product input states that
 are the eigenstates of the first $3$ Pauli operators and the output qubits are measured in a product basis of eigenstates
of the last $3$ Pauli operators. A careful analysis reveals that all average values of tensor products including $\sigma_I$ can be determined from
measurements of other average values where a different Pauli operator replaces $\sigma_{I}$.
Even after this reduction we are left with $63$ average values that need to be determined.
 This method thus requires $63\times 8= 504$ different measurement settings,
which is $21$ times higher than the $24$ measurement settings required by our protocol.

In our experiment we have access to only $4$ out of the $8$ possible
measurement outcomes, because part of the signal
is deflected by the second calcite beam displacer and not detected. Moreover, in order to avoid the need to calibrate
the detector efficiencies, we have utilized only a single coincidence signal D2$\&$D3. For each measurement
setting we used $8$ different waveplate settings such that the coincidence detection of photons
by detectors D2 and D3 corresponded to the projection onto $8$ different eigenstates of the $3$-qubit measurement basis.
Estimation of fidelity using the procedure of Refs. \cite{Flammia11,Silva11,Steffen12} would thus in our case require $8\times 504=4032$ different settings.
Just preparing all these settings would take about $17$ hours which exceeds the stability time of the Mach Zehnder interferometer formed by the calcite 
beam displacers and of the Hong-Ou-Mandel dip on the central partially polarizing beam splitter.
By contrast, our fidelity estimation procedure requires only $8 \times 24=192$ different settings
even if we use only a single coincidence signal D2$\&$D3. For each waveplate setting we measured the number of coincidences in $100\, {\rm s}$  and
for each input state we detected about $6.6\times 10^4$ coincidences in total.
The whole measurement lasted about $6$ hours and most of this time was used for data acquisition.

\begin{figure}[!t!]

\centerline{\includegraphics[width=0.99\linewidth]{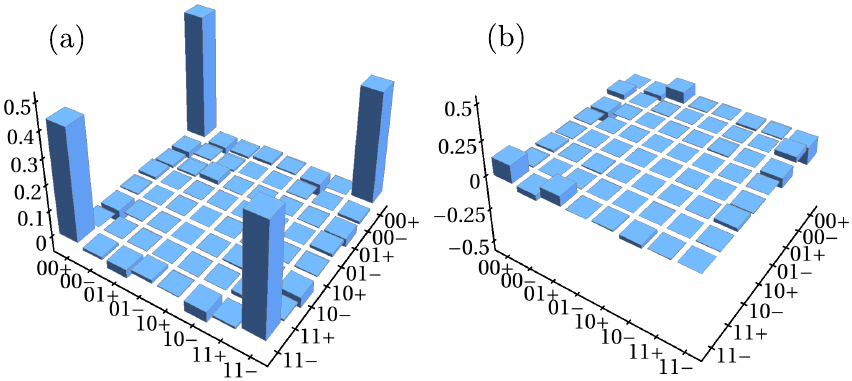}}

\caption{Real (a) and imaginary (b) parts of reconstructed density matrix of a three-qubit state generated by the CCZ gate from input state $|\Phi_2^{+}\rangle|+\rangle$.}
\end{figure}

At a cost of larger number of required measurement settings, Monte Carlo sampling allows to obtain an estimate of the true gate fidelity
 while our approach yields only a lower and upper bound on the gate fidelity.
 One may conjecture that with increasing number of qubits the Monte Carlo sampling
will eventually become more efficient even in terms of the number of experimental settings,
because the number of required distinct settings depends only on the desired
accuracy of the Monte Carlo estimate and not on the system size  \cite{Flammia11,Silva11}. Let us therefore discuss this issue in more detail.
If one wants to obtain a Monte Carlo fidelity estimate $F_E$
that with probability $p$ will be $\epsilon$-close to the true fidelity $F$, $\mathrm{Pr}(|F_E-F|>\epsilon)\leq 1-p$,
then the number of required experimental settings reads
\begin{equation}
M\approx\frac{1}{(1-p)\epsilon^2}.
\label{M}
\end{equation}
These $M$ settings need to be randomly chosen from all the relevant settings according to a specific relevance distribution \cite{Flammia11,Silva11}.
According to formula (\ref{M}),  achieving a $1\%$ precision ($\epsilon=0.01$) with $90\%$ probability ($p=0.9$) would require about $M=10^5$ settings.
For a three-qubit CCZ gate this largely exceeds the total number of relevant settings.
In such case one can directly perform measurements for all the relevant settings \cite{Steffen12}, which eliminates
the error of fidelity estimate due to Monte Carlo sampling. As shown in the previous section, the $N$-qubit controlled-Z gate becomes exponentially close
to the identity operation with the increasing number of qubits, c.f. Eq. (\ref{FIformula}).  In order to reliably certify that the gate fidelity exceeds $F_\mathrm{I}$,
the error $\epsilon$ has to be of the order of $2^{-N}$ for large $N$. For this specific task we thus obtain asymptotic scaling  $M \propto 2^{2N}$.
In contrast, our method requires only $N 2^N$ distinct measurement settings.

We can conclude that the gate fidelity estimation based on the lower bound (\ref{FboundN}) is suitable for systems where a change of measurement setting
requires significant effort. On the other hand, Monte Carlo sampling is particularly  well suited for platforms where
changes of measurement settings are fast and easy to implement. Both techniques could also be combined. For instance, our method can be used for a 
quick assessment of the gate performance during its construction and fine-tuning, followed by a more precise but also more time-consuming characterization of the resulting gate by Monte Carlo sampling.

\subsection{Generation of three-qubit GHZ state}

In our experiment, entanglement between spatial and polarization qubits of the signal photon can be deterministically generated and controlled
by a suitable rotation of the HWP1 that addresses only the lower path in the interferometer formed by the two beam displacers.
This allows us to investigate the action of the three-qubit CCZ gate on
partially entangled input states. In particular, we have tested the fusion of a two-qubit maximally entangled Bell state
\begin{equation}
|\Phi_2^{+}\rangle=\frac{1}{\sqrt{2}}(|00\rangle+|11\rangle)
\end{equation}
with a third qubit prepared in state $|+\rangle=\frac{1}{\sqrt{2}}(|0\rangle+|1\rangle)$ into a three-qubit maximally entangled GHZ state,
\begin{equation}
U_{\mathrm{CCZ}}|\Phi_2^{+}\rangle|+\rangle= \frac{1}{\sqrt{2}}(|00+\rangle+|11-\rangle).
\label{GHZ}
\end{equation}
 The output state was reconstructed with the Maximum Likelihood method and it is shown in Fig. S.1. Note that a Hadamard basis $|+\rangle,|-\rangle$ is used for the third qubit in this figure.
  The generated state is very close to the ideal GHZ state (\ref{GHZ}) as witnessed by a high state fidelity $F=0.962$ and  purity $\mathcal{P}=0.949$.
Note the nonzero imaginary parts of the off-diagonal density matrix elements which can be attributed to the uncompensated residual phase shift in the
interferometer, $\phi_0\approx \pi/11$. If we compensate for this phase shift on the experimental data, the fidelity increases to $\tilde{F}=0.972$.


\begin{thebibliography}{99}
\bibitem{SI}
See Appendix for detailed proof of the bound (\ref{FboundNmain}), comparison of our technique with Monte Carlo sampling, and discussion of
 generation of a maximally entangled three-qubit GHZ state.
\end{thebibliography}
\end{document}